# Cleaning ability of mixed solutions of sulfonated fatty acid methyl esters

Veronika I. Yavrukova,[a] Dimitar N. Shandurkov,[a] Krastanka G. Marinova,[a]

Peter A. Kralchevsky,[a,]* Yee Wei Ung,[b] Jordan T. Petkov[b,1]

[a] *Department of Chemical and Pharmaceutical Engineering, Faculty of Chemistry and Pharmacy, Sofia University, 1164 Sofia, Bulgaria*

[b] *KLK OLEO, KL-Kepong Oleomas Sdn Bhd, Menara KLK, Jalan PJU 7/6, Mutiara Damansara, 47810 Petaling Jaya, Selangor Dalur Ehsan, Malaysia*

———————————

\* Corresponding author: Tel. +359-2-8161262; *E-mail address*: pk@lcpe.uni-sofia.bg

[1] Present address: Arch UK Biocides Ltd., Hexagon Tower, Crumpsall Vale, Blackley, Manchester M9 8GQ, UK

**Abstract**

Here, we present results from a systematic study on cleaning of oily deposits from solid surfaces (porcelain and stainless steel) by solutions of fatty acid sulfonated methyl esters (SME), sodium salts. The zwitterionic dodecyldimethylamine oxide (DDAO) has been used as a cosurfactant. As representatives of the vegetable and mineral oils, sunflower seed oil and light mineral oil have been used. The process of oil drop detachment from the solid substrates (roll-up mechanism) has been monitored. In the case of porcelain, excellent cleaning of oil is achieved by mixed solutions of SME and DDAO. In the case of stainless steel, excellent cleaning (superior than that by linear alkylbenzene sulfonate and sodium lauryl ether sulfate) is provided by binary and ternary mixtures of SMEs, which may contain also DDAO. For the studied systems, the good cleaning correlates neither with the oil/water interfacial tension, nor with the surfactant chainlength and headgroup type. The data imply that governing factors might be the thickness and morphology of admicelle layers formed on the solid/water interface. The results indicate that the SME mixtures represent a promising system for formulations in house-hold detergency, having in mind also other useful properties of SME, such as biodegradability, skin compatibility and hard water tolerance.

# 1. Introduction

Sulfonated methyl esters (SME) of fatty acids are subject to increasing interest during the last two decades. SMEs are derived from renewable sources and are considered as a green alternative of petroleum-derived surfactants (Siwayanan et al., 2014; Lim et al., 2016, 2019; Jin et al., 2016; Maurad et al., 2017; Xiu et al., 2017). SME are insensitive to the water hardness, unlike the linear alkylbenzene sulfonates, which are widely used in cleaning formulations (Cohen et al., 1999, Lim et al., 2016, 2019; Ivanova et al., 2017; Xiu et al., 2017). Mixtures of SME and linear alkylbenzene sulfonates (LAS) have been shown to improve the LAS solubility in hard water and have been used to achieve significant builder's reduction in detergent formulations (Lim et al., 2019). Foaminess and foam stability are better with SME as compared to LAS (Lim et al., 2016; Tai et al., 2018). Detergency and cleaning by SME and SME+LAS formulations have been found comparable and even better than those with LAS alone (Maurad et al., 2017; Lim et al., 2019; Tai et al., 2018).

Along with the detergency characterization, there is a considerable advance in the physicochemical characterization and theoretical modeling of adsorption and micellization of the different SMEs and their mixtures with other ionic or nonionic surfactants (Patil et al., 2004; Wong et al., 2012; Danov et al., 2015; Ivanova et al., 2017; Basheva et al., 2019, Xu et al., 2018; Wang et al., 2019).

Recent studies describe direct measurements of the cleaning performance of different SME in specialized tests evaluating the detergency power of powders (Siwayanan et al., 2014; Lim et al., 2019), laundry liquids (Maurad et al., 2017) or dishwashing liquids (Tai et al., 2018). These studies are helpful for industry to select the most suitable surfactants for commercial detergent products. However, the mechanism of soil removal processes has not been investigated using the available methods for characterizing the soil detachment at nano-, micro- and macro-scales (see e.g. Cuckston et al., 2019).

In the literature, there are few results about the factors that govern the cleaning performance. Indications that the soil detachment can depend on the surfactant alkyl chainlength (Gambogi et al., 2006; Siwayanan et al., 2014, Lim et al., 2016) and headgroup nature (Mahdi et al. 2015) have been found. The oil/water interfacial tension is also an important factor for oil drop detachment (Phaodee et al., 2018). Another governing factor is expected to be the surface energy of the solid substrate, which significantly depends on the surface type, treatment, ageing, etc. (see e.g. Hedberg et al., 2014; Kim et al., 2016, Tsujii, 2017).

The detergency action of anionic surfactant solutions could be improved by the addition of zwitterionic cosurfactant, which has been used in various formulations such as shampoos, hand and body washes for foam boosting (Basheva et al., 2000); cleaning aids (Gambogi et al., 2006), and rheological thickeners (Christov et al. 2004). Zwitterionic surfactants have been found to improve also the solubilization capacity of the ionic surfactants (Golemanov et al. 2008). Interactions in mixed solutions of zwitterionic and anionic surfactants have been described by different approaches and synergistic effects have been found (Hines et al., 1998; Danov et al., 2004; Angarska et al., 2004; Basheva et al., 2019). Amine-oxide surfactants are among the most widely used zwitterionics and special attention has been paid to their salt and pH sensitivity (Maeda et al., 1995, 1996; Singh et al., 2006; Schellmann et al., 2015). To the best of our knowledge, so far there is no study that relates the surface and bulk properties of SME solutions and of SME+zwitterionic mixtures to their action as detergents.

Our goal in the present article is to investigate the cleaning of oily deposits from solid surfaces by SME solutions and by SME+zwitterionic mixtures. As zwitterionic cosurfactant, dodecyldimethylamine oxide (DDAO) is used. The investigated solid substrates are porcelain and stainless steel, which are typical materials for kitchenware. As representatives of the vegetable and mineral oils, sunflower seed oil (SFO) and light mineral oil (LMO) have been used. The cleaning efficacy is characterized by direct monitoring of the process of oil drop detachment from the substrates in the investigated surfactant solutions; see e.g. Rowe et al., 2002; Kolev et al., 2003, Kralchevsky et al., 2005, and Davis et al., 2006.

Section 2 describes the used materials and methods. In Section 3, the investigated systems are characterized by contact angles; interfacial tensions and Krafft temperature of the surfactants. In Section 4, the results from the experiments on oil drop detachment are reported and discussed with respect to the factors, which govern the rather different cleaning performance of surfactants that have very similar chemical nature. We believe the results would be of interest to both industrial researchers developing new formulations and academic scientists investigating the physicochemical mechanisms of detergency.

## 2. Materials and methods

*2.1. Materials*

The used sulfonated methyl esters, SME ($\alpha$-sulfo fatty acid methyl ester sulfonates, sodium salts, denoted also $\alpha$-MES), are products of the Malaysian Palm Oil Board (MPOB) and KLK OLEO.

In particular, C12-SME, C14-SME, and C16-SME are sulfonated methyl esters of the respective fatty acids: lauric, myristic, and palmitic. C1618-SME, represents a mixture of 85 wt% palmitic (C16-SME) and 15 wt% stearic (C18-SME) sulfonated methyl esters and has a mean molecular weight $M$ = 376.70 g/mol and critical micellization concentration (CMC) = 0.9 mM. C1618-SME is preferred in applications because of its lower Kraft temperature and better water solubility (Schambil & Schwuger, 1990). The surfactant samples were used as received.

We used also linear alkyl benzene sulfonate, sodium salt (LAS) product of Sigma Aldrich; sodium laurylethersulfate with two ethylene oxide groups (SLES) product of KLK OLEO, and N,N-Dimethyldodecylamine N-oxide (DDAO), product of Sigma Aldrich. The transition from the cationic to the zwitterionic form of DDAO occurs near pH = 6 (Maeda et al., 1995; Schellmann et al., 2015).

As inorganic salt additives we used sodium chloride, NaCl (Honeywell, Germany) and calcium dichloride hexahydrate, $CaCl_2 \cdot 6H_2O$ (Sigma Aldrich). To adjust the desired pH of the solutions, HCl or NaOH were used. The solutions were prepared with deionized water from Elix 3 (Millipore) water purification system.

All experiments were performed with solutions of 0.2 and 0.5 wt% total surfactant concentrations, which are typical in cleaning applications (Jin et al., 2016, Lim et al., 2019). The solutions were prepared by intensive stirring for 1 to 24 hours prior use. The solutions of C14-, C16- and C1618-SME were heated to 40 °C for a better and faster dissolution. The pH was adjusted after the full surfactant dissolution.

Salt concentration is known to affect significantly the surface and interfacial tension of ionic surfactant solutions (Kralchevsky et al., 1999, 2002; Fainerman & Luccassen-Reinders 2002, Gurkov et al., 2005). To fix the ionic strength of the studied solutions, NaCl was added at molar concentration that is ca. five times higher than the total molar surfactant concentration. Thus, all solutions with 0.5 wt% total surfactant concentration contain 0.365 wt% (≈62 mM) NaCl, whereas all solutions with 0.2 wt% total surfactant concentration proportionally contain 0.146 wt% (≈25 mM) NaCl.

As model liquid soils, we used sunflower seed oil (SFO) and light mineral oil (LMO). Food grade sunflower oil was purchased from a local supplier and used after purification by passing through a column filled with the absorbents Silicagel 60 (Fluka, cat. # 60741) and Florisil® (60/100 mesh, Supelco, cat. # 20280-U). LMO product of Sigma-Aldrich (cat # 33,077-9) was used as received.

As solid substrates, we used glaze porcelain and stainless steel. Rectangular porcelain plates of size 3×3×0.5 cm were cut from a white feldspar porcelain dinnerware plates. The plates were cleaned by soaking in ethanol, abundant rinsing with deionized water and drying at ambient temperature. Stainless steel AISI 304 (Cold rolled, bright annealed, average roughness 0.05 – 0.1 μm) was used as rectangular flat plates of size 2×2×0.1 cm. The stainless steel sample plates were cleaned by consecutive soaking in ethanol and in Decon 90$^{TM}$ liquid detergent, abundant rinsing with deionized water, and drying at ambient temperature for several hours.

*2.2. Experimental methods*

The solutions' surface and interfacial tensions, $\sigma_{AW}$ and $\sigma_{OW}$, were determined by using the "pendant /buoyant drop" method. For this goal, a buoyant bubble or drop was formed on the tip of a J-shaped hollow needle dipped in the aqueous solution. The surface tension was determined by drop shape analysis (Rotenberg et al. 1983; Hoorfar & Neumann 2006) with the software DSA1 on the instrument DSA10 (Krüss GmbH, Hamburg, Germany).

To determine precisely the Krafft temperature for 0.5 wt% surfactant solutions, we measured the solutions' turbidity using a UV/VIS spectrophotometer Jasco V-700 at wavelength 500 nm. The temperature was decreased by 0.5° steps starting from 30 °C. The samples were tempered for 10 minutes at each temperature and the measurements were performed afterwards. The Krafft temperature was determined from the onset of rise of turbidity (Heckmann et al., 1987; Tzocheva et al., 2012).

Contact angles of water and oil drops on the used solid surfaces were determined by side observations using the instrument DSA10 and DSA1 software (Krüss GmbH, Germany).

*2.3. Monitoring the oil drop detachment in surfactant solutions*

The systematic observation of soil removal has been realized by using a glass cuvette mounted on the instrument DSA10. The procedure is as follows. First, we put a dry substrate on the bottom of the 50 ml rectangular glass cuvette. Next, 3 μl oil drop is placed on the substrate and its three-phase contact angle, $\theta$, is measured (see Fig. 1A). The drop is left at rest for 10 min.

Afterwards, 20 ml surfactant solution is gently poured in the cuvette. The shrinking of the oil-drop/substrate contact area (with possible drop detachment) has been observed for 15 min (Fig. 1B), and the contact angle variation has been recorded. All experiments have been performed at least twice using at least two separate substrates of each type for each solution and oil.

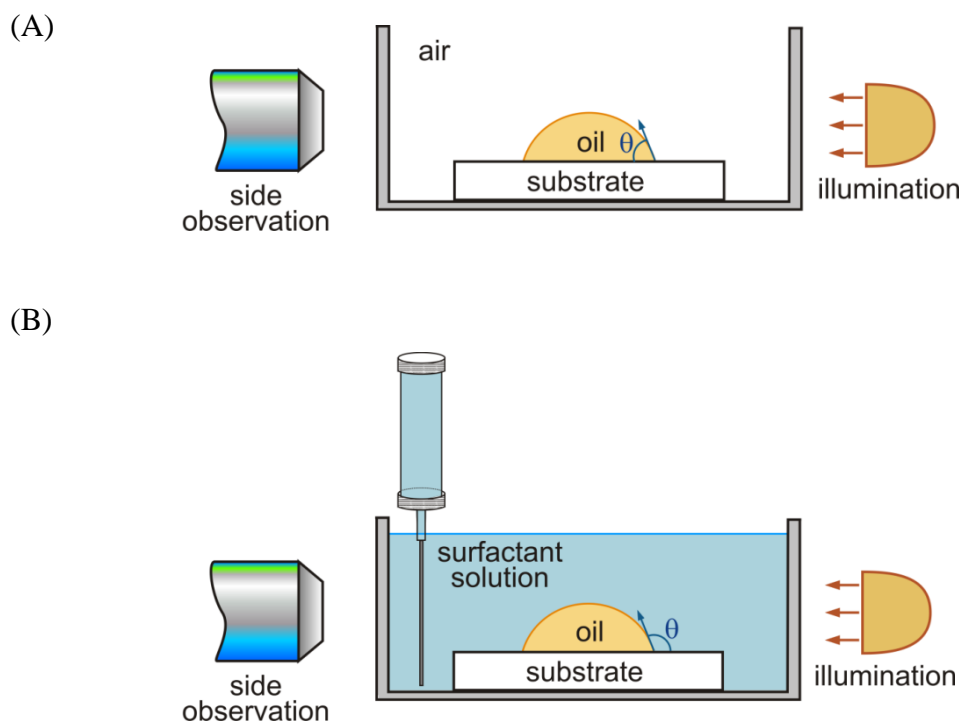

**Fig. 1** Schematic presentation of the setup used to measure contact angles. (A) Drop on a solid substrate in air; the contact angle $\theta$ is measured across the *liquid* phase. (B) Drop on a solid substrate in aqueous solution; the contact angle $\theta$ is measured across the *water* phase.

After the aqueous phase is poured in the experimental cell (Figure 1B), the contact area oil/substrate begins to shrink. At that, we distinguish three scenarios of oil drop evolution (Figure 2A).

(i) *No detachment*: The shrinkage of the three-phase contact line (and the decrease of contact angle $\theta$) decelerates and stops at a relatively large contact angle, e.g. $\theta > 80°$ (Figure 2B). The drop remains attached to the substrate.

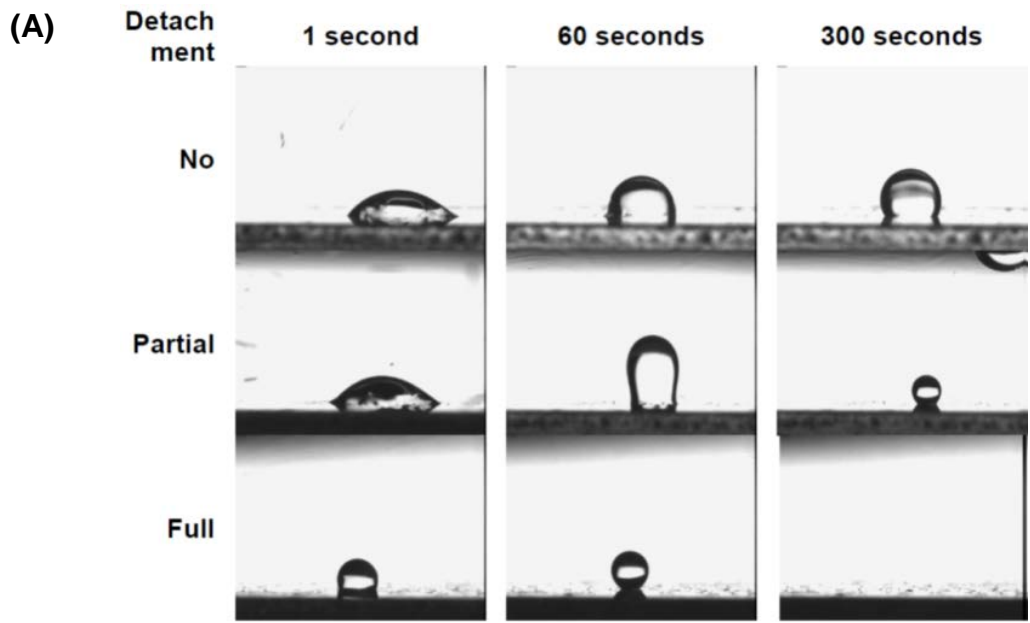

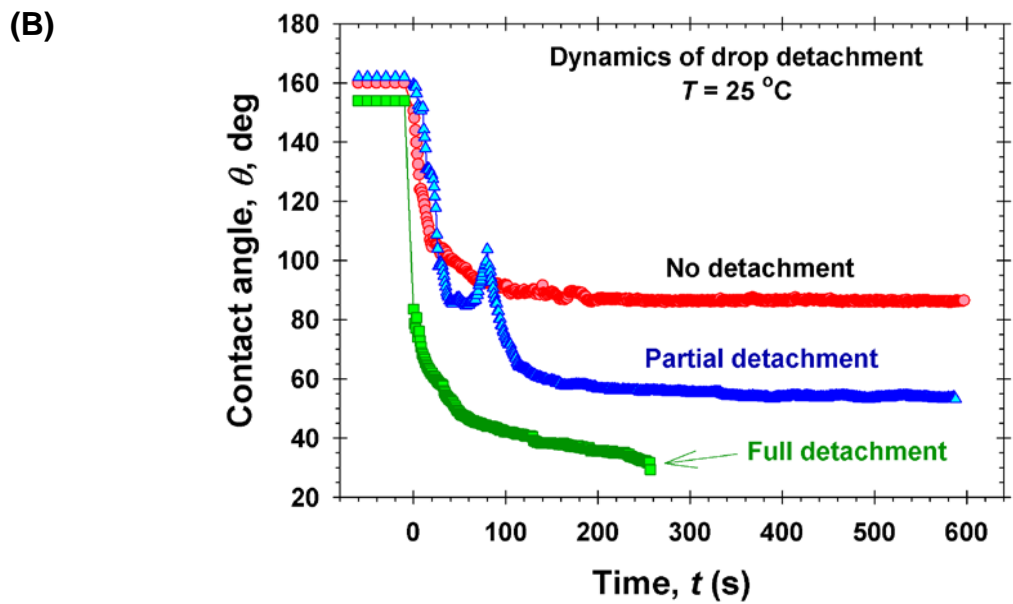

**Fig. 2** (A) Photographs illustrating the three scenarios of evolution of oil drops in surfactant solutions: (i) No detachment; (ii) Partial detachment and (iii) Full detachment. (B) The respective typical variations of contact angle $\theta$. The moment $t = 0$ corresponds to pouring the surfactant solution in the experimental cell.

(ii) *Partial detachment*: At a certain stage of the contact-line shrinkage, necking instability appears and the drop breaks to two parts at the neck. The upper (larger) part is detached, whereas the lower part remains fixed to the substrate as a residual drop. In Figure 2B, the moment of drop breakage corresponds to the local maximum of $\theta$.

(iii) *Full detachment*: In this case, the necking instability occurs within few seconds after pouring the surfactant solution. After that, the contact angle of the residual droplet gradually decreases up to its full detachment (Figure 2). Such behavior corresponds to the roll-up mechanism of cleaning (see e.g. Tsujii, 1998; Smulders, 2002; Shi et al., 2006).

In some experiments, a residual droplet remains attached to the substrate with a small contact angle (across water), but this drop detaches if it is subjected to a minor external force (which is present in real cleaning experiments). Such a case will be referred as *almost full detachment*.

## 3. Experimental characterization of the studied systems

### 3.1. Contact angle measurements

The cleaned dry porcelain and stainless steel substrates were characterized by measuring the drop contact angle $\theta$ of 3 µl drops of water, SFO and LMO; the upper phase is air (Fig. 1A). The results are shown in Table 1, where the values of $\theta$ are average over at least 6 different drops on more than 3 different plates of the same material. The standard deviation is ±5°.

The data for water drops show that porcelain is markedly more hydrophilic that stainless steel. This leads to much easier cleaning of oily soils from porcelain than from stainless steel (see below).

The comparison of the data for SFO and LMO indicates that the mineral oil wets the solid substrate much better than the vegetable oil. This fact is related to the use of mineral oils as lubricants. However, both oils wet the solid surfaces better than water, which can be explained with the greater contribution of dispersion interaction to the surface free energy in the case of oils (Israelachvili 2011).

**Table 1** Solid/liquid/air contact angle, $\theta$, for drops of deionized water, sunflower oil (SFO), and light mineral oil (LMO); $T = 25$ °C.

| Substrate | Solid/liquid/air contact angle, $\theta$ | | |
|---|---|---|---|
| | Water | SFO | LMO |
| Porcelain | 38° | 23° | <10° |
| Stainless steel | 53° | 18° | <10° |

*3.2. Solutions' surface tension and turbidity*

We performed measurements of the surface tension of 0.2 wt% solutions of the used SME surfactants in the absence and presence of DDAO. In the solutions with DDAO, the weight fractions of SME and DDAO are, respectively, 0.8 and 0.2, the total surfactant concentration being the same, viz. 0.2 wt%. The pH was varied in the range between 4 and 8. It should be noted that the dishwashing liquids in the market have pH in the range from 6 to 10 (see e.g. Shi et al., 2006) but the requirement for skin mildness gives preference to formulations with pH close to 6. The range $4 < \text{pH} < 8$ is used also in skin cleansing formulations (Gambogi et al., 2006; Harmalker & Lai, 2006).

Figure 3 shows data for the pH dependence of the surface tension, $\sigma_{AW}$, of the studied solutions. The most significant effect in this figure is the lowering with 8–10 mN/m of $\sigma_{AW}$ for C14-SME and C16-SME solutions upon the replacement of a part (20 %) of SME with DDAO. This effect is expected to favor the cleaning of oils from solid surfaces. Similar surface tension drop has been observed for SDS-DDAO mixtures (Angarska et al., 2004) and has been practically applied for optimization of dish washing formulations (Shi et al., 2006).

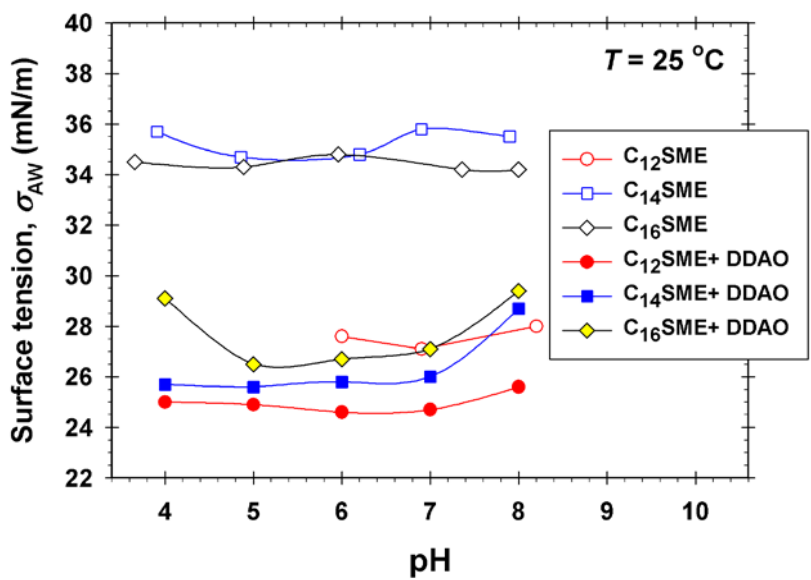

**Fig. 3** Effect of DDAO on the surface tension of SME solutions, $\sigma_{AW}$, at various pH. For all solutions, the total surfactant concentration is 0.2 wt% and they contain also 0.146 wt% NaCl. The weight fractions of SME and DDAO in the mixed solutions (full symbols) are, respectively, 0.8 and 0.2.

For C12-SME (without DDAO) $\sigma_{AW}$ is markedly lower as compared to C14- and C16-SME. This can be explained with the presence of admixture of unsulfonated lauric-acid methyl ester (a residual component from the synthesis) in the used surfactant sample. The replacement of a part of C12-SME with DDAO further lowers $\sigma_{AW}$, which takes the lowest values among the solutions characterized in Figure 3.

The data in Figure 3 do not show any strong effect of pH on $\sigma_{AW}$ in the investigated concentration range. A shallow minimum of $\sigma_{AW}$ is observed only for C16-SME + DDAO. In our subsequent experiments, pH = 6 is fixed, which is close to the aforementioned minimum.

In relation to the influence of water hardness, we studied the effect of added $CaCl_2$ on $\sigma_{AW}$ (Table 2). The concentration of added $Ca^{2+}$ was 5 mM, which corresponds to very hard water. The addition of 5 mM $Ca^{2+}$ decreases $\sigma_{AW}$ with 1–2 mN/m. The effect is stronger for SLES as compared to C16-SME, in agreement with the finding for relatively low binding energy of the $Ca^{2+}$ ions to the sulfonate groups of SME (Ivanova et al. 2017). The lowering of $\sigma_{AW}$ with 7–9 mN/m due to DDAO is a much stronger effect than that of 5 mM $Ca^{2+}$.

**Table 2** Surface tension, $\sigma_{AW}$, of solutions at total surfactant concentration 0.5 wt% with 0.365 wt% added NaCl, in the presence or absence of DDAO and $CaCl_2$; pH = 6 and $T$ = 25 °C.

| Surfactant | $Ca^{2+}$ (mM) | $\sigma_{AW}$ (mN/m) |
|---|---|---|
| SLES | 0 | 31.7 |
| SLES | 5 | 29.4 |
| 4:1 SLES/DDAO (w/w) | 5 | 24.5 |
| C16-SME | 0 | 34.6 |
| C16-SME | 5 | 31.2 |
| 4:1 C16-SME/DDAO (w/w) | 0 | 26.7 |
| 4:1 C16-SME/DDAO (w/w) | 5 | 25.8 |

Unlike the clear solutions of SLES and C16-SME with $Ca^{2+}$ presented in Table 2, the solutions of 0.5 wt% LAS + 5 mM $Ca^{2+}$ were very turbid. This is related to the high sensitivity of LAS to hard water. However, if the concentrations of surfactant and calcium are decreased to 0.2 wt% and 0.9 mM, respectively, all studied solutions become clear (Fig. 4A).

Furthermore, to visualize the sensitivity of LAS to $Ca^{2+}$ we increased the surfactant and calcium concentrations to 0.5 wt% and 2.25 mM, respectively (Figure 4B). As expected, the solution of LAS is the most turbid. The 4:1 LAS/DDAO solution is less turbid, which is due to the replacement of a part of LAS with DDAO. The solutions of SLES and C16-SME are completely clear. However, the presence of DDAO in the solutions of SLES and C16-SME slightly increases the turbidity (Figure 4B). We could hypothesize that $Ca^{2+}$ is able to bridge between the zwitterionic form of DDAO and two anionic surfactant molecules, which leads to precipitation of the formed hydrophobic complex with three alkyl chains.

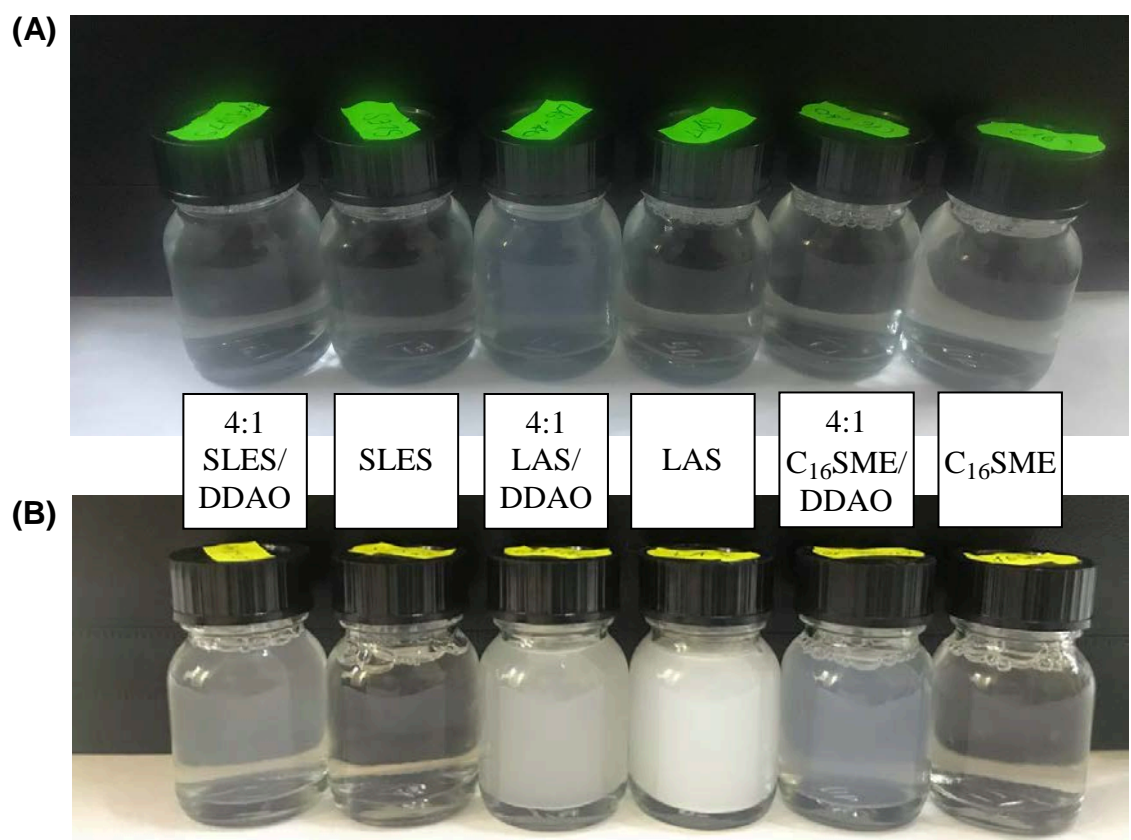

**Fig. 4** Photos of glass vials containing (A) 0.2 wt% total surfactant concentration + 0.9 mM $Ca^{2+}$; (B) 0.5 wt% total surfactant concentration + 2.25 mM $Ca^{2+}$ (hard water). In both cases pH = 6 and $T = 25$ °C. The mix ratios are by weight (w/w).

By turbidimetry, we measured the Krafft temperature, $T_K$, of some of the studied solutions, which are clear at 25 °C. Among the systems in Table 3, C14-SME has the lowest $T_K = 10.1$ °C, because of its shortest alkyl chain. In contrast, C16-SME has the highest $T_K$, which is slightly below 25 °C. In agreement with literature evidence (Schambil & Schwuger, 1990), C1618-SME has a significantly lower Krafft temperature, $T_K = 18.0$ °C. Another binary surfactant mixture, 4:1 C16-SME /C12-SME, has $T_K$ = 22.0 °C, which is lower than that of C16-SME alone, but higher than that of C1618-SME, despite the shorter chain of C12-SME. Finally, a ternary surfactant mixture, 3:1:1 C16-SME /C12-SME /DDAO, has Krafft temperature $T_K$ = 17.9 °C, which is close to that of C1618-SME.

**Table 3** Krafft temperature of unary, binary and ternary surfactant solutions at a total surfactant concentration of 0.5 wt% containing also 0.365 wt% NaCl at pH=6. The mix ratios are by weight.

| Surfactant | Krafft temperature |
|---|---|
| C14-SME | 10.1 °C |
| C16-SME | 24.8 °C |
| C1618-SME | 18.0 °C |
| 4:1 C16-SME /C12-SME | 22.0 °C |
| 3:1:1 C16-SME /C12-SME /DDAO | 17.9 °C |

**4. Results from the cleaning experiments and discussion**

*4.1. Cleaning of oils by solutions of single surfactant*

Systematic study on oily soil removal with drops from SFO and LMO deposited on porcelain and stainless steel was carried out with six ionic surfactants: LAS, SLES, C12-, C14-, C16- and C1618-SME (see Figure 5). All cleaning aqueous solutions contained 0.5 wt% surfactant and 0.365 wt% NaCl at pH = 6.

In the case of *porcelain* substrate, the results are as follows. For the vegetable oil (SFO), with all six studied surfactants we observed *full detachment* of the drops within less than a minute. In contrast, for the mineral oil (LMO) with all six surfactants we observed *no detachment* of the drops – both cases are illustrated in Figure 2. This difference correlates with the fact that LMO wets better porcelain than SFO; see Table 1.

In the case of *stainless steel* substrate, the results are presented in Figure 5A. These results are rather surprising. With LAS, SLES, C14-SME and C16-SME *no detachment* of oil drops is observed. In contrast, *full detachment* of both SFO and LMO drops is observed with C1618-SME. With C12-SME, we observed full detachment of the SFO drops and partial detachment of the LMO drops.

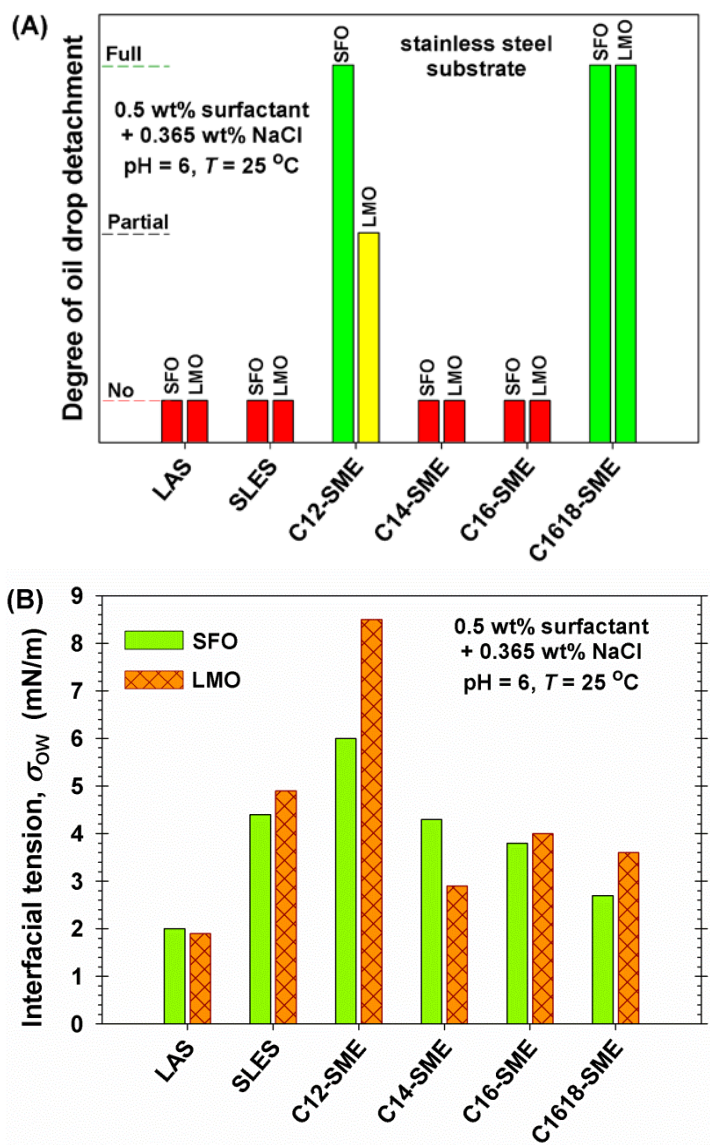

**Fig. 5** (A) Results from the drop detachment experiments with SFO and LMO on stainless steel in solutions of anionic surfactants. (B) The respective values of oil/water interfacial tension, $\sigma_{OW}$.

The comparison of Figures 5A and 5B indicates that there is no correlation between low interfacial tension and good cleaning performance. Indeed, the LAS solution has the lowest $\sigma_{OW}$, whereas the C12-SME solution – the highest one. However, the cleaning performance of C12-SME is much better than that of LAS.

Likewise, the comparison of Figure 5A with the data in Table 3 shows that there is no correlation between low Krafft temperature and good cleaning. Indeed, C14-SME has lower Krafft temperature than C1618-SME. However, the cleaning performance of C1618-SME is much better than that of C14-SME.

There is no correlation also between the surfactant chainlength and the cleaning performance. Indeed, C12-SME and C1618-SME have different chainlengths, but similar cleaning performance (Figure 5A). Moreover, the chainlengths of LAS, SLES and C12-SME are similar, but their cleaning performance is rather different.

In addition, surfactants with identical headgroups, C12-, C14-, C16- and C1618-SME exhibit very different cleaning performance (Figure 5A). This means that both the headgroups and hydrocarbon chains of surfactant molecules matter for the cleaning process. This fact implies that the formation of surfactant adsorption bilayers or admicelles on the solid surface affect the detachment of oil drops. The morphology of the admicellar layer seems to be very specific and dependent on the kind of solid surface and surfactant type; see e.g. Zhang & Somasundaran, 2006; De Oliveira Wanderley Neto et al., 2014; Atkin et al., 2001; Wangchareansak et al., 2013. We could hypothesize that admicelles of appropriate morphology can penetrate in the wedge-shaped region near the three-phase contact line and can act as "molecular jacks" that promote the full detachment of the oil drops from the substrate.

*4.2. Mixed solutions of anionic surfactant and DDAO*

To improve the cleaning action of the surfactant solutions, we added the zwitterionic surfactant DDAO to the solutions of SLES, C12-, C14-, C16- and C1618-SME. In this series of experiments, the ratio anionic/zwitterionic surfactant was 4:1 (w/w) and the total surfactant concentration was 0.5 wt%. For both SFO and LMO, the presence of DDAO lowered the oil/water interfacial tension $\sigma_{OW}$ below 1 mM/m. (The used DSA method does not allow one to measure precisely interfacial tensions lower than ca. 1 mN/m; see e.g. Hoorfar & Neumann, 2006). Despite the fact that the interfacial tension was < 1 mN/m, it was not low enough to cause spontaneous emulsification, so that the mechanism of drop removal was roll-up again.

In the case of *porcelain* substrate, the treatment with the aforementioned 4:1 anionic/DDAO surfactant solutions leads to *full detachment* of the drops from both SFO and LMO; see Figure 2. In other words, the presence of DDAO very essentially improves the cleaning of LMO from porcelain – see Section 4.1.

In the case of *stainless steel* substrate, the results are presented in Figure 6A. The SLES+DDAO solution provides partial detachment of both SFO and LMO drops. The treatment with C12-SME + DDAO solution leads to full detachment of SFO drops, but partial detachment of LMO drops. For C14-SME + DDAO solutions, the roles are exchanged – partial detachment of SFO, but full detachment of LMO. The treatment with C16-SME + DDAO solution leads to full detachment of SFO drops, but no detachment of LMO drops. Finally, with and without DDAO the solutions of C1618-SME provide full detachment of both SFO and LMO drops; see Figures 5A and 6A.

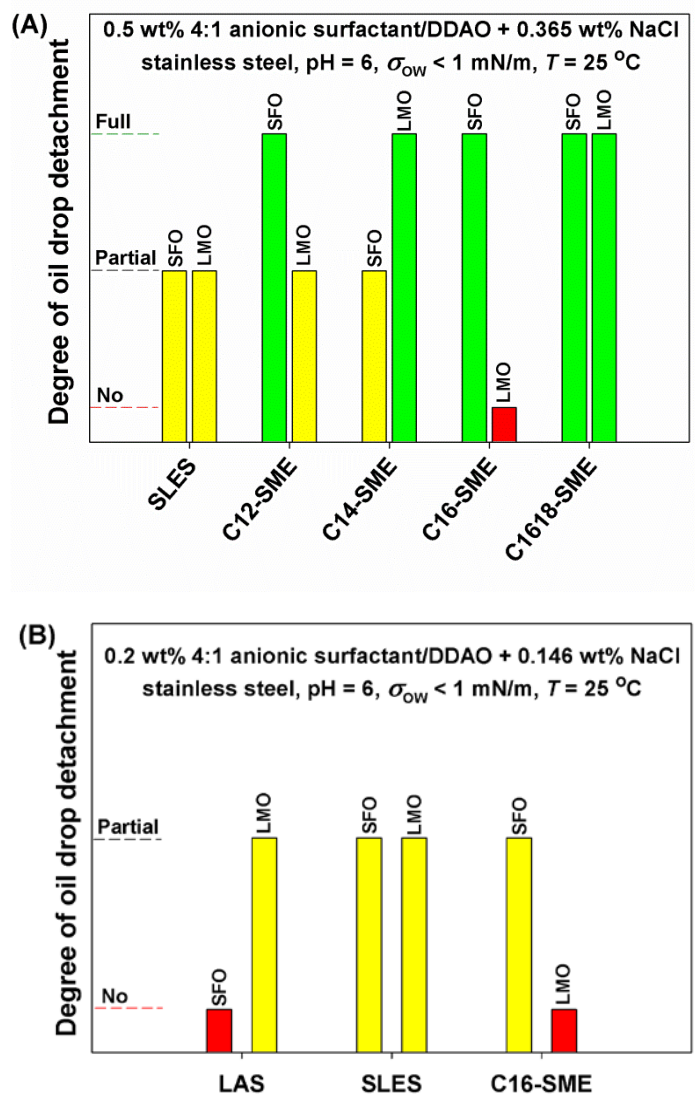

**Fig. 6** (A) Results from the drop detachment experiments with SFO and LMO on stainless steel in 4:1 (w/w) solutions of anionic surfactant and DDAO. (A) 0.5 wt% total surfactant concentration + 0.365 wt% NaCl; (B) 0.2 wt% total surfactant concentration + 0.146 wt% NaCl.

The solutions with 0.5 wt% 4:1 LAS/DDAO are turbid. To avoid the precipitation, we decreased the total surfactant concentration to 0.2 wt%. In Figure 6B, we compare the cleaning performance of 0.2 wt% 4:1 LAS/DDAO solutions with the performance of 0.2 wt% 4:1 SLES/DDAO solutions and 0.2 wt% 4:1 C16-SME/DDAO solutions for oil drops on stainless steel. The lowering of the total surfactant concentration from 0.5 to 0.2 wt% does not affect the cleaning performance of SLES solutions – at both concentrations we observe partial detachment of the oil drops; see Figures 6A and 6B. However, the lowering of the total surfactant concentration worsens the cleaning of SFO by C16-SME solutions – from full detachment to partial detachment. Finally, the treatment with LAS + DDAO solution leads to full detachment of LMO drops, but no detachment of SFO drops.

In summary, the presence of DDAO in the surfactant solutions markedly improves the cleaning of oily deposits from stainless steel; see Figures 5A and 6A. In general, the decrease of the total surfactant concentration from 0.5 to 0.2 wt% worsens the cleaning performance, compare Figures 6A and 6B. This is not surprising, because the average thickness and morphology of the adsorbed surfactant (admicelles) on the solid surface are expected to essentially depend on surfactant concentration (Zhang & Somasundaran, 2006; De Oliveira Wanderley Neto et al., 2014).

Mixing of amine-oxide surfactants with anionic surfactants is known to boost the growth of wormlike micelles in the bulk of solution (Hoffmann et al., 1992). In particular, the mixing of SME with the zwitterionic surfactant cocamidopropyl betaine (with or without added electrolyte) produces a strong synergistic effect on the micelle growth in the bulk (Yavrukova et al., 2020). The present results on cleaning indicate that the mixing of SME with the zwitterionic DDAO could promote also the formation of admicelles on the surface of stainless steel.

*4.3. Cleaning of oils by binary SME solutions*

Because all studied 0.5 wt% solutions of anionic surfactants + DDAO lead to full detachment of SFO and LMO drops from porcelain (Section 4.2), our investigations have been continued with stainless steel substrates, which are more difficult to clean. We recall that no detachment of oil drops has been observed in solutions of C14-SME and C16-SME (Figure 5A).

Here, we investigate whether the mixing of C14-SME and C16-SME with shorter chain SMEs could improve the cleaning of oily stains. For this goal, we monitored the detachment of SFO and LMO drops from stainless steel substrates in mixed solutions of 0.5 wt% total surfactant concentration and composition 4:1 C14-SME/C12-SME; 4:1 C16-SME/C12-SME, and 4:1 C16-SME/C14-SME.

The results are presented in Figure 7A. The best results (full detachment of both SFO and LMO drops) were obtained with the 4:1 C16-SME/C14-SME solutions, which performs similarly to C1618-SME – compare the rightmost columns in Figures 5A and 7A. The treatment with the other two mixed solutions, 4:1 C14-SME/C12-SME and 4:1 C16-SME/C12-SME, leads to full detachment of LMO drops (which is a considerable improvement), but no detachment of SFO drops was observed.

The comparison of the drop detachment data in Figure 7A with the respective data for the interfacial tension $\sigma_{OW}$ in Figure 7B shows the absence of any correlation again. In such a case, the different behaviors of the studied surfactant solutions should be related to the three-phase contact angle $\theta$ (Fig. 1B) that, in turns, depends on the solid/water interfacial tension, $\sigma_{SW}$ (Tsujii, 2017). As already mentioned, the values of $\sigma_{SW}$ are affected by the structure and morphology of the surfactant adsorption layers on the solid/water interface, which may include formation of admicelles.

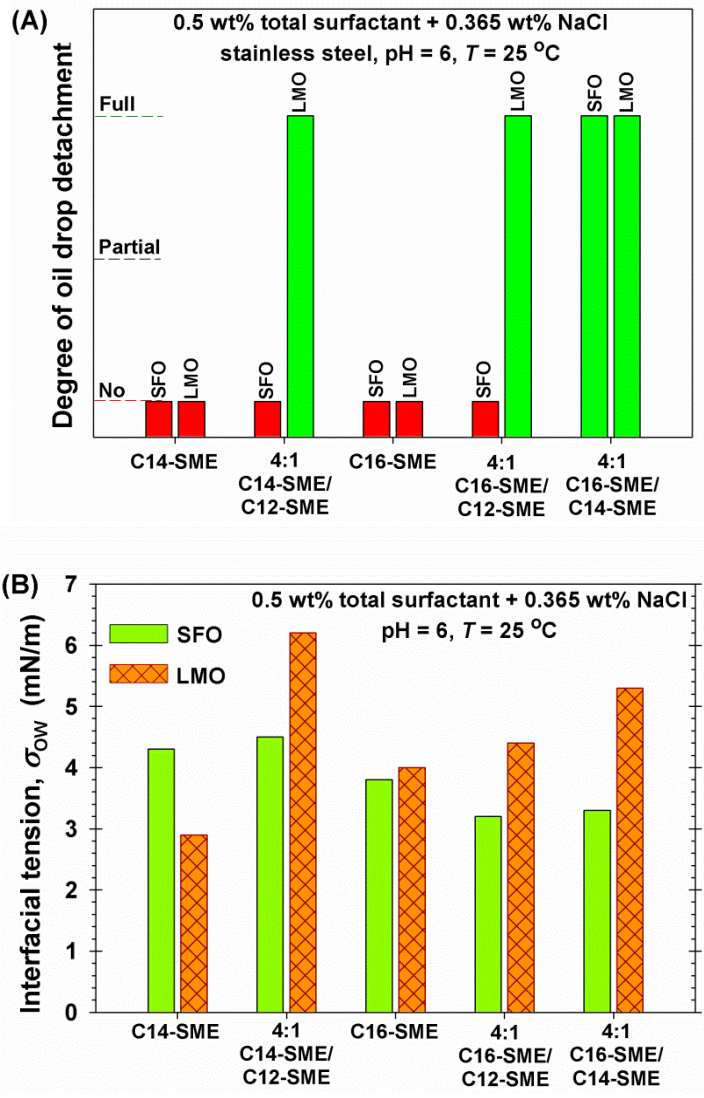

**Fig. 7** Comparison of data for unary and binary SME solutions. (A) Results from the drop detachment experiments with SFO and LMO on stainless steel. (B) The respective values of oil/water interfacial tension, $\sigma_{OW}$. The mix ratios are by weight (w/w).

*4.4. Cleaning of oils by ternary surfactant solutions with DDAO*

Finally, we added the zwitterionic surfactant DDAO to the double mixtures from Figure 7A to verify whether further improvement of oil cleaning from stainless steel could be achieved. The results are shown in Figure 8. In the presence of DDAO, the excellent cleaning performance of the C16-SME + C14-SME mixture is preserved. Moreover, the presence of DDAO improves the cleaning performance of the C14-SME + C12-SME and C16-SME + C12-SME mixtures – for both SFO and LMO we observe almost full detachment of the oil drops. This means that the main mass of the oil drop has been detached and only a small (nano-liter) drop has remained on the substrate. Such small drop can be easily detached under the action of a minor mechanical force.

The mixing of anionic SMEs with a zwitterionic surfactant gives rise to the growth of giant wormlike micelles in the bulk of surfactant solution (Basheva et al., 2019; Yavrukova et al., 2020). The results in Figure 8 could be an indication that the mixing of SMEs with DDAO promotes the formation of mixed surfactant layers, possibly – admicelles, on the surface of stainless steel. Mixed solutions of amine oxide with alkyl sulfates or alkyl ethoxy sulfates find wide applications in grease cleaning (Lant & Keuleers, 2016). The results in Figures 7 and 8 show that superior cleaning could be obtained by combinations of SME surfactants, with or without amine oxide, thus, avoiding the use of LAS (hard-water sensitive) as in other studies (Maurad et al., 2017; Lim et al., 2019; Tai et al., 2018).

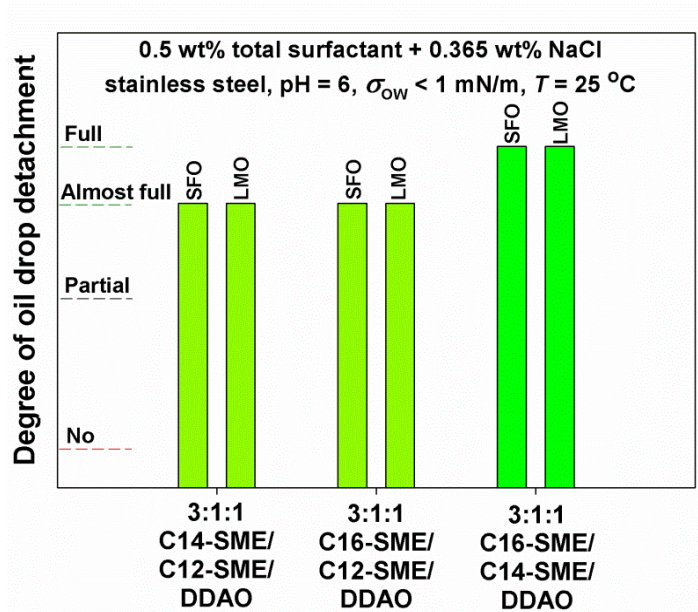

**Fig. 8** Results from the drop detachment experiments with SFO and LMO on stainless steel in ternary mixed solutions of SMEs and DDAO. The mix ratios are by weight.

## 4. Conclusions

The present article reports results from a systematic study on the cleaning of oily deposits from porcelain and stainless steel by solutions of sulfonated methyl esters (SME) of fatty acids, sodium salts. As a cosurfactant, dodecyldimethylamine oxide (DDAO) has been used. Comparative experiments with LAS and SLES have been also performed. As representatives of the vegetable and mineral oils, sunflower seed oil and light mineral oil have been used. The process of oil drop detachment from the solid substrates (roll-up mechanism) was studied by direct observations (Figures 1 and 2). In general, the surfactants are expected to promote the detachment of oil drops from the substrate by lowering the oil/water and solid/water interfacial tensions, $\sigma_{OW}$ and $\sigma_{SW}$ (Tsujii, 2017). In view of potential applications in house-hold detergency, all experiments have been carried out at pH = 6 (mild to skin).

The experiments showed that excellent cleaning of oil from porcelain can be achieved by the mixed solutions of SME and DDAO (Section 4.2). For this reason, all subsequent experiments were focused on cleaning of oil from stainless steel.

In the case of single surfactant solutions, full oil drop detachment from stainless steel was observed with C1618-SME and C12-SME, whereas no drop detachment was observed for C14-SME, C16-SME, LAS and SLES (Figure 5A). The addition of DDAO improves the cleaning by C14-SME and C16-SME, but only with respect to one of the two types of oil (Figure 6A). The mixing of C14-SME and C16-SME leads to excellent cleaning performance (Figure 7A), which is similar to that of C1618-SME. Finally, excellent cleaning was obtained also with ternary surfactant solutions, composed of two SMEs and DDAO (Figure 8).

The results for the investigated systems indicate that the good cleaning of oils from stainless steel correlates neither with the oil/water interfacial tension, nor with the surfactant chainlength, headgroup type, or Krafft point (Section 4.1; Figures 5 and 7). The only possible explanation remains the lowering of $\sigma_{SW}$ that could be caused by formation of admicelles on the solid surface (Zhang & Somasundaran 2006; De Oliveira Wanderley Neto et al, 2014). The cleaning action seems to be influenced by the morphology of the admicellar layer, which is very specific and depends on the kind of solid surface and surfactant mixture (Atkin et al., 2001; Wangchareansak et al., 2013). This issue could be a subject of subsequent studies, where the formation of admicelles could be confirmed by appropriate experimental methods, e.g., atomic force microscopy (AFM) or appropriate spectroscopy methods.

The results show that binary and ternary mixtures of SMEs, which may contain also DDAO, exhibit excellent cleaning performance (superior than that of LAS and SLES) for the two types of oils and two types of substrates. For this reason, the SME mixtures represent a promising system for formulations in house-hold detergency, having in mind also other useful properties of SME, such as biodegradability, skin compatibility and high hard water tolerance.

**Acknowledgments**

All authors gratefully acknowledge the support from KLK OLEO. KM and PK acknowledge the support from the Operational Programme "Science and Education for Smart Growth", Bulgaria, grant number BG05M2OP001-1.002-0023. VY acknowledges the financial support received from the program "Young Scientists and Postdoctoral Candidates" of the Bulgarian Ministry of Education and Science, MCD № 577/17.08.2018. The authors are grateful to Dr. Hui Xu and Dr. Emily Tan from KLK OLEO, who consulted the work on project.